\documentstyle[12pt,aaspp4,epsf]{article}

\begin{document}

\newcommand{\twcott}{$^{12}$CO $J$=3$-$2}
\newcommand{\twcoto}{$^{12}$CO $J$=2$-$1}
\newcommand{\twcooz}{$^{12}$CO $J$=1$-$0}
\newcommand{\twco}{$^{12}$CO}
\newcommand{\thcott}{$^{13}$CO $J$=3$-$2}
\newcommand{\thcoto}{$^{13}$CO $J$=2$-$1}
\newcommand{\thco}{$^{13}$CO}
\newcommand{\jtt}{$J$=3$-$2}
\newcommand{\jto}{$J$=2$-$1}
\newcommand{\joz}{$J$=1$-$0}
\newcommand{\etal}{{\it et al.}}
\newcommand{\eg}{{\it e.g.}}
\newcommand{\ie}{{\it i.e.}}

\title{The Effect of Star Formation on Molecular Clouds in Dwarf Irregular
Galaxies: IC 10 and NGC 6822} 
\author{Glen R. Petitpas and Christine D. Wilson}
\affil{Department of Physics and Astronomy}
\affil{McMaster University, 1280 Main Street West, Hamilton Ontario, Canada L8S
4M1}
\authoremail{petitpa@impatiens.physics.mcmaster.ca}

\begin{abstract}

We have observed the \twcoto\ and \jtt\ lines at a few locations in the
dwarf irregular galaxies IC 10 and NGC 6822 using the James Clerk
Maxwell Telescope. In addition, we have observed the \thcoto\ line for
IC 10 and the first detection of the \thcott\ transition in a Local
Group galaxy. The CO line ratios in IC 10 are uniform and are
consistent with the average line ratios observed in M33 at the
$1\sigma$ level. These low metallicity environments appear to be porous
to UV radiation and allow for more efficient heating of molecular gas
by nearby \ion{H}{2} regions. The \twcott/\jto\ ratio for the molecular
cloud in NGC 6822 is higher than those found for IC 10 and M33 and
suggests that the \twco\ emission is optically thin in this region.
This high line ratio is likely the result of its location inside a
large \ion{H}{2} region with low metallicity and low gas content, and
requires a hydrogen density greater than $10^4$ cm$^{-3}$ and a kinetic
temperature greater than 100 K. The \twco/\thco\ \jtt\ line ratio in
one of the molecular clouds in IC 10 indicates that the gas must have a
rather high kinetic temperature of about 100 K. In IC 10 we observe
structures on a variety of size scales that all appear to be
gravitationally bound. This effect may help explain the rather high
star formation rate in IC 10.

\end{abstract}

\keywords{galaxies: irregular --- galaxies: individual (IC 10, NGC 6822) ---
galaxies: ISM --- ISM: molecules --- Local Group}

\section{Introduction}

Dwarf irregular galaxies are rather simple systems, free from the
complicated dynamics that accompany bars or spiral density waves. This
simplicity makes them useful for studying the interactions between star
formation and the interstellar medium (ISM). Temperature, density, and
abundance may all affect the star formation process, and thus, determining
the physical conditions inside molecular clouds is important for
understanding how the properties of the molecular gas affect the type and
amount of stars formed. Unfortunately, extragalactic molecular clouds
have low filling factors even within beams as small as 15$''$ and so
their true temperatures cannot be deduced from the observed peak
temperature of the spectral line. Instead, observations of many
rotational transitions of CO and its isotopomers can be combined with
radiative transfer models to constrain the physical conditions inside
these clouds. As it is necessary to observe some of the rarer isotopomers
such as \thco, these techniques have been primarily applied to starburst
galaxies (\eg\ \markcite{til91}Tilanus \etal\ 1991). 

 Relatively few dwarf galaxies have been detected in CO and the lines
that are observed are much weaker than those found in spiral galaxies
(\markcite{com86}Combes 1986; \markcite{tac87}Tacconi \& Young 1987;
\markcite{arn88}Arnault \etal\ 1988). The weakness of these lines is
likely due to a lower CO emissivity per unit mass of gas relative to
spiral galaxies, which is caused by the low metallicity of most
irregular galaxies (12 + log(O/H) = 8.16 for IC 10, 8.20 for NGC 6822,
\markcite{leq79}Lequeux \etal\ 1979). IC 10 is one of the few dwarf
galaxies in which strong CO lines have been detected
(\markcite{bec90}Becker 1990), with line strengths that are greater
than those of the LMC, SMC, and NGC 6822 (\eg\ \markcite{isr93}Israel
\etal\ 1993; \markcite{wil94}Wilson 1994; \markcite{isr97}Israel
1997). The simplest explanation of these strong CO lines is that IC 10
contains large amounts of molecular gas (\markcite{wil91}Wilson \& Reid
1991). Unfortunately, except for the strong CO lines, there is no
additional evidence to support this explanation. In fact, IC 10 has a
rather low gas-to-dust ratio for a dwarf irregular galaxy
(\markcite{sch93}Schmidt \& Boller 1993). A more likely answer lies in
the star forming activity. \markcite{mas95}Massey \& Armandroff (1995)
found that IC 10 has a {\it global} surface density of Wolf-Rayet stars
three times as high as any other Local Group galaxy, comparable to the
surface density found within active star forming regions of M33. This
high concentration of Wolf-Rayet stars reinforces the idea that IC 10
is undergoing a rather strong burst of star formation. It has been
shown both observationally and theoretically that CO emission depends
on the conditions in the molecular clouds. \markcite{mal88}Maloney \&
Black (1988) find that the ratio of CO intensity to excitation
temperature is constant for ensembles of identical clouds. Hence,
higher temperatures will result in brighter CO emission. The high star
formation rates could increase the temperature of the molecular gas and
more readily excite the CO molecules. Such a process would be able to
produce the strong CO lines observed in IC 10.

In this paper we study the physical properties of molecular clouds in
environments with different amounts of star forming activity located in
two Local Group dwarf irregular galaxies, IC 10 and NGC 6822. IC 10
appears optically as a patchy nebulous region extending approximately $3'
\times 4'$. It has been classified as a dwarf irregular of the Magellanic
type (IBm; \markcite{dev65}de Vaucouleurs \& Ables 1965). It is rather
close to the Galactic plane (b $\sim -3\arcdeg$), which indicates that
there is heavy obscuration at visual wavelengths. It is considered a
Local Group member based on its distance of 0.82 Mpc
(\markcite{wil96}Wilson \etal\ 1996). The bright \ion{H}{2} regions (\eg\
\markcite{san74}Sandage \& Tammann 1974) indicate that IC 10 is
undergoing massive star formation. NGC 6822 is also a dwarf irregular
galaxy relatively close to the Galactic plane (b $\sim -18\arcdeg$). It
has a bar-like appearance extending approximately $10' \times 20'$
north-south with a prominent ``wing'' located to the southeast. Its
distance of 0.49 Mpc (\markcite{mcg83}McGonegal \etal\ 1983) also places
it in the Local Group. In \S\ref{obs} we discuss the observations and the
data reduction process. In \S\ref{results} we discuss the results found
for the individual molecular clouds in IC 10 and NGC 6822 based on the
line ratios and models. We also compare masses obtained from flux
measurements with virial masses and discuss the validity of the
assumptions made to aid in the interpretation of the data. In
\S\ref{alpha} we discuss the implications of these new data and the
physical conditions of the molecular clouds for the CO-to-H$_2$
conversion factor. The paper is summarized in \S\ref{concl}.

\section{Observations and Data Reduction \label{obs}}

Four regions centered on giant molecular clouds in IC 10 were observed
using the James Clerk Maxwell Telescope (JCMT) over the period of 1994
October 20 -- 24. The regions were selected from both interferometric
CO \joz\ surveys (\markcite{wil91}Wilson \& Reid 1991;
\markcite{wil95}Wilson 1995) and a single dish \twcoto\ and
\joz\ survey (\markcite{bec90}Becker 1990). All four of the regions
were observed in \twcoto\ with follow-up observations for 2 of these
regions (containing clouds IC 10:MC 1/2 \& IC 10:MC 6/7/8) in
\twcott\ and \thcoto. A final \thcott\ observation was obtained at a
single position located at the peak of the \thcoto\ emission in IC10:MC
1/2. The half-power beamwidth of the JCMT is $22''$ at 230 GHz
(\twcoto) and $15''$ at 345 GHz (\twcott). Data were obtained on a
five-point cross offset by $11''$ for the \jto\ data and $8''$ for the
\jtt\ data. One giant molecular cloud in NGC 6822 was observed during
service observing at the JCMT on the nights of 1996 May 5 (\twcott) and
1994 June 8 (\twcoto). The cloud was observed at a single position in
the \jto\ transition and a five-point cross offset by $8''$ in the
\jtt\ transition. All observations were obtained using the Dwingeloo
Autocorrelation Spectrometer.

The data were reduced using the Bell Labs data reduction package COMB.
The data were binned to 1.5 km s$^{-1}$ resolution and first to seventh
degree baselines were removed. The lines were quite narrow ($<$ 30 km
s$^{-1}$) and the spectrometer bandwidth was at least 450 km s$^{-1}$,
so the removal of rather high degree baselines for some spectra do not
introduce significant errors into the final line intensities. Third and
seventh degree baseline removal on the same data yielded line
parameters that agreed to within 1$\sigma$. The \twcott\ data were then
convolved to simulate a $22''$ beam. The spectra for each cloud are
shown in Figures \ref{spectra} and \ref{MC12spectra}. The data were
then binned to the same resolution and shifted to the same $V_{\rm
LSR}$ (when necessary). This allowed the calculation of line ratios for
each channel within the spectral line. The ratios for each channel were
averaged to obtain the final line ratios. Only those channels with a
signal-to-noise ratio of at least two in the line ratio were used in
the calculation. This method helps reduce errors introduced by the
existence of poor baselines or weak signals. The uncertainty in the
value for the line ratio is taken as the standard deviation of the
mean. As a final step, the line ratios were scaled to the main beam
temperature scale using the appropriate values for $\eta_{\rm MB}$ (see
below). A summary of the line ratios is given in Table \ref{ratios} and
a channel by channel histogram of the ratios is shown in Figure
\ref{3ratios}.

\placefigure{spectra}
\placefigure{MC12spectra}
\placetable{ratios}
\placefigure{3ratios}

The calibration of the IC 10 data was monitored by frequently observing
both planets and spectral line calibrators. Unfortunately, it was later
determined that the beam of the telescope was decidedly non-circular
during 1994 October (see JCMT Newsletter No. 4, March 1995), enough to
affect our calibration. Although the spectral line calibrators showed
considerable scatter, with individual measurements differing by as much
as $\pm 40\%$ from standard spectra, observations of a given source
were repeatable to within $\sim$5\%. The wide scatter of the spectral
line calibrators may be understood given the large ellipticity of the
beam and the range in diameter of the calibrators, from compact carbon
stars to more extended emission in \ion{H}{2} regions. Depending on the
size of the source, the measured line strength may be larger or smaller
with the elliptical beam compared to that measured with a circular
beam. The main beam efficiencies determined using Saturn were $0.60 \pm
0.02$ at 220/230 GHz and $0.36\pm 0.01$ at 345 GHz. Both efficiencies
are substantially lower than normal values. Since Saturn had a
diameter of 18$''$ during these observations, it provides a reasonable
match to the size of the emission regions in IC 10. This assumption was
confirmed by observing the molecular cloud MC 20 in M33, for which
observations obtained with the JCMT in a more normal state were
available; the \twcoto\ line from this source was indeed low by about
20\%. Thus, we adopt these observed efficiencies to convert our data to
the main beam temperature scale. For the NGC 6822 data (taken at a
different dates under normal calibration conditions), observations of
CRL 2688 agreed with the published values, and so we adopt the normal
main beam efficiencies from the JCMT Users Guide of 0.72 at 230 GHz and
0.58 at 345 GHz.

Since the line strengths we are comparing are measured with the same
beam diameter, using the $T_{\rm R}^*$ temperature scale would ensure
that our observed line ratios are equal to the true radiation
temperature ratios. However, conversion to $T_{\rm R}^*$ from $T_{\rm
A}^*$ requires knowledge of the forward scattering and spillover
($\eta_{\rm FSS}$), which is difficult to measure and was not attempted
during the observing run. As a result, we do not know how to correct
the published values of $\eta_{\rm FSS}$ to take into account the
calibration problems discussed above. On the other hand, we do have
good values for the main beam efficiencies and so an accurate
conversion to main beam temperature ($T_{\rm MB}$) is possible. Using
the $T_{\rm MB}$ scale instead of $T_{\rm R}^*$ scale only changes the
\twcott/\jto\ line ratio by $\sim$5\% under normal calibration
conditions. For this reason, we will use the main beam temperature
scale for the line ratios throughout this paper.

\section{Molecular Cloud Properties \label{results}}

\subsection{MC 1/2 and MC 6/7/8 in IC 10 \label{MC12.68}}

 The clouds MC 1/2 are located 60$''$ (240 pc at a distance of 0.82 Mpc)
from the center of the brightest \ion{H}{2} region in IC 10 (no.~111,
\markcite{hod90}Hodge \& Lee 1990; \markcite{sho89}Shostak \& Skillman
1989). This \ion{H}{2} region has an H$\alpha$ luminosity of $4 \times
10^{38}$ erg s$^{-1}$. Conversely, MC 6/7/8 is not located near a large
\ion{H}{2} region; it is $86''$ (340 pc) away from the edge of even the
nearest moderate sized one (no.~50, H$\alpha$ luminosity of $\sim
10^{37}$ erg s$^{-1}$, \markcite{hod90}Hodge \& Lee 1990). For these two
regions (MC 1/2 and MC 6/7/8), the \twco/\thcoto\ ratios agree very
well, which suggests similar \twco/\thco\ abundances in these clouds.
Although the \twcott/\jto\ ratios also agree within the large
uncertainties, the larger ratio of 0.90 is found for the cloud nearer
the bright \ion{H}{2} region, while the lower ratio of 0.66 is found in
the more quiescent cloud.

 \markcite{wil97}Wilson, Walker, \& Thornley (1997) found that the
 giant molecular cloud NGC 604-2, located in the very intense star
formation environment of the giant \ion{H}{2} region NGC 604 in M33,
had a \twcott/\jto\ ratio that was higher than average for M33
($1.07\pm 0.17$ compared to $0.64\pm 0.07$ for clouds without optical
\ion{H}{2} regions). They also found that a cloud located only 120 pc
($31''$) away from NGC 604 did not show an increase in the
\twcott/\jto\ ratio. In IC 10, MC 1/2 does exhibit an enhanced
\twcott/\jto\ line ratio, even though it is further away from the
weaker \ion{H}{2} region no.~111. This result suggests that if a
similar line ratio enhancement is taking place as in M33, the ISM in
the metal-poor dwarf irregular galaxy IC 10 must be more porous to UV
radiation. This effect could allow the line ratio enhancement to occur
at greater distances from weaker \ion{H}{2} regions than is possible in
the higher metallicity spiral galaxy M33 (12 + log(O/H) = 8.48 in NGC 604, \markcite{vil88}Vilchez \etal\ 1988).

 The IC 10 data and the NGC 6822 data (see \S\ref{MC2}) support the
conclusions of \markcite{wil97}Wilson \etal\ (1997) that the
\twcott/\jto\ line ratio increases towards the centers of bright
\ion{H}{2} regions. The higher transitions can be more readily excited
by higher temperatures or higher densities, either of which will
produce higher line ratios. It is believed that higher temperatures are
present near star-forming regions, but without accurate densities for
these clouds, we cannot assess the extent to which both of these
factors contribute to the elevated line ratios. Independent temperature
information on these clouds (\eg\ from the new bolometer array SCUBA on
the JCMT) should yield a clearer picture of the processes at work.
 
 For the clouds for which we have observed more than one line ratio, we
can use models to help constrain the physical conditions of the clouds.
We will use the large velocity gradient (LVG) model, which assumes that
the clouds in the beam have internal velocities large enough to Doppler
shift the CO emission from one part of the cloud sufficiently to prevent
reabsorption by other CO molecules in other parts of the cloud
(\markcite{sco74}Scoville \& Solomon 1974). We used the LVG code RAD written
by Dr.~L.~Mundy and implemented as part of the MIRIAD data reduction
package. Model grids were run for kinetic temperatures $T_{\rm K}$ = 10,
15, 20, 30, 50, 100, 200, 300 K, abundance ratios [\twco]/[\thco] = 30,
50, 70, and covering a range of densities of log$[n_{\rm
H_2}(\rm{cm}^{-3})]$ = 1 -- 6 and \twco\ column densities per unit
velocity of log$[N($\twco$)/\Delta v~~ (\rm{cm^{-2}~km^{-1}~s})]$ = 15 --
20 . To calculate the \twco\ column densities from $N/\Delta v$, we use
the full-width half maximum velocity. 
 
 Figure \ref{MC12lvg} shows one of the two solutions found for the line
ratios of IC 10:MC 1/2. The three ratios obtained using the JCMT
(\twco/\thcott, \twcott/\jto\ and \twco/\thcoto) are shown as bands that
indicate the $1\sigma$ upper and lower limits of the line ratio.
Solutions are indicated by the locations where the regions of the
different ratio boundaries overlap. Of the 24 possible models, this
only occurred twice for the JCMT line ratios of
IC 10:MC 1/2, at $T_{\rm K}$ = 100 K and for the abundances
[\twco]/[\thco] = 50 and 70. Both cases indicate a density of $10^4 -
10^5$ cm$^{-3}$ and column densities per unit velocity of about
$10^{17.7}$ cm$^{-2}$ (km s$^{-1})^{-1}$. The average full-width half
maximum of the emission lines in IC 10:MC 1/2 is 12.2 $\pm$ 1.4 km
s$^{-1}$, which gives a true column density of $6 \times 10^{18}$
cm$^{-2}$.

\placefigure{MC12lvg}

We have also calculated the \twcoto/\joz\ ratio by combining our JCMT
\twcoto\ data ($22''$ beam) and the IRAM 30-m telescope \twcooz\ data
($21''$ beam) obtained by \markcite{bec90}Becker (1990). The similarity
in the beam sizes at these two frequencies allows a straightforward
combination of the data sets without the use of convolution algorithms.
In an attempt to constrain the abundance better, we then included the
\twcoto/\joz\ line ratio into the LVG model. However, instead of
overlapping with one of the other two existing solutions, the
\twcoto/\joz\ line ratio provided two additional solutions for each
abundance at lower density ({\it n}(H$_2$) = $10^{1.5} - 10^{2.5}$
cm$^{-3}$). There is no region where all four line ratios overlap.
This result may indicate the existence of separate high and low density
regions that contribute to the emission. The high density region
dominates the upper $J$ emission while the lower $J$ emission would
originate predominantly from the lower density region (as would be
predicted by the Boltzmann equation). We know that Galactic molecular
clouds are clumpy (\eg\ \markcite{stu90}Stutzki \& G\"usten 1990), and
there is every reason to believe that extragalactic molecular clouds
are clumpy as well. If this is the case we can perhaps interpret the
two distinct emitting regions as representing the high density clumps
and the less dense inter-clump regions. Unfortunately, the lack of data
with resolution better than 7\arcsec or 28 pc (\markcite{wil91}Wilson \&
Reid 1991; \markcite{oht92}Ohta \etal\ 1992; \markcite{wil95}Wilson 1995)
precludes further discussion of the degree of substructure within molecular
clouds in IC 10.

With only two line ratios (\twcott/\jto\ and \twco/\thcoto), it is not
possible to put very strong constraints on the physical conditions in IC
10 MC 6/7/8 without making some assumptions. For example, if we assume a
moderate temperature of 20 K and a [\twco]/[\thco] abundance of 50 (as
found for IC 10:MC 1/2), we can estimate upper and lower limits for the
densities in the cloud. The results of this model are shown in Figure
\ref{MC68lvg}. There is a rather broad range of allowed values for density
($10^{1.6} - 10^{4}$ cm$^{-3}$) and for \twco\ column densities per unit
velocity ($10^{16.5} - 10^{18.5}$ cm$^{-2}$ (km s$^{-1})^{-1}$). The
full-width half maximum of the emission lines in MC 6/7/8 is 7.1 $\pm$ 1.6
km s$^{-1}$, which gives true column densities ranging from $2 \times
10^{17}$ to $2 \times 10^{19}$ cm$^{-2}$. If we assume a [\twco]/[\thco]
abundance of 70 (as is also allowed by the more complete IC 10:MC 1/2 model)
the only change is that the lower limit on column density increases
from $2 \times 10^{17}$ cm$^{-2}$ to $7 \times 10^{17}$ cm$^{-2}$,
narrowing the allowed range of column density slightly. The range in
density and the upper limit in column density remain essentially the
same. Adopting a higher kinetic temperature of 30 K results only in a
slightly lower upper limit on H$_2$ density (10$^{3.6}$ cm$^{-3}$
instead of 10$^{4}$ cm$^{-3}$). Adopting a cooler kinetic temperature
of 10 K indicates that H$_{2}$ densities greater than 10$^{4}$ cm$^{-3}$
are required to produce the observed \twcott/\jto\ and \twco/\thcoto\
line ratios. As in IC 10:MC 1/2, the \twcoto/\joz\ line ratio for MC 6/7/8 does
not overlap with the JCMT line ratios, which again suggests we may be
seeing a combination of high and low density gas in this region. 

\placefigure{MC68lvg}

\subsection{MC 2 in NGC 6822 \label{MC2}}

The molecular cloud MC 2 in NGC 6822 is located almost in the very
center (within $1''$) of the brightest \ion{H}{2} region in NGC 6822.
Discovered by \markcite{hub25}Hubble (1925), and designated \ion{H}{5}
by \markcite{hod88}Hodge \etal\ (1988), this \ion{H}{2} region has an
H$\alpha$ luminosity of $4 \times 10^{38}$ erg s$^{-1}$
(\markcite{hod89}Hodge, Lee \& Kennicutt 1989). The \twcott/\jto\ ratio
(1.40 $\pm$ 0.07) is significantly higher than either of the IC 10 line
ratios. The \ion{H}{2} region \ion{H}{5} is weaker than the giant
\ion{H}{2} region NGC 604 in M33 (H$\alpha$ luminosity of $5 \times
10^{39}$ erg s$^{-1}$, \markcite{ken88}Kennicutt 1988) yet we see a
much higher \twcott/\jto\ line ratio in NGC 6822:MC 2. Again, as in IC
10, we see that in the low metallicity environment of NGC 6822,
molecular clouds associated with \ion{H}{2} regions can attain line
ratios higher than in higher metallicity galaxies such as M33.

 High $J$ transitions can be readily excited by high temperatures, high
densities, or some combination of the two. We would expect the
temperature of molecular clouds near \ion{H}{2} regions to be higher
than those unassociated with \ion{H}{2} regions. These higher
temperatures would more readily excite the higher $J$ transitions even
in relatively low density regions. In the low density, hot gas there
will be less CO in the lowest rotational levels and hence a lower
probability of being reabsorbed (\ie\ lower optical depth). Without
separate constraints on the density, it is difficult to assess the role
that density plays in producing this high line ratio. However, it can
be shown from the virial theorem that for clouds with a relatively
constant surface pressure the product of density and temperature
($\rho~ T$) is approximately constant (\eg\ equation (2.8d),
\markcite{mcl96}McLaughlin \& Pudritz 1996). This result implies that
for virialized clouds, an increase in temperature results in a decrease
in density. It is likely that this cloud in NGC 6822 is in virial
equilibrium (Wilson 1994, and next section). It also seems likely in
the case of MC 2 that the temperature should be higher than in typical
interstellar molecular clouds because of its close proximity to the OB
stars generating the large \ion{H}{2} region. We therefore conclude
that in MC 2 the high line ratio is most likely the result of the high
temperatures due to the proximity of the bright \ion{H}{2} region.

With only one line ratio, a detailed LVG analysis is not possible. If
we assume the cloud is in local thermodynamic equilibrium (LTE), we
would expect line ratios no greater than unity for optically thick gas,
while optically thin gas can produce much larger ratios
(\eg\ \markcite{wal91}Walker 1991). The observed line ratio indicates
that the \twco\ gas is optically thin in MC 2. CO is the most abundant
molecule after H$_2$ and is generally found to be optically thick in
giant molecular clouds. Optically thin \twco\ is quite rare, and likely
the result of special physical conditions inside an \ion{H}{2} region
or possibly in starburst galaxies.

For gas to be molecular, it must be shielded by dust or have a high
enough column density to be self-shielding. For molecular gas to
collapse into stars, it is believed that the column density must exceed
some critical value (\eg\ \markcite{mou76}Mouschovias \& Spitzer 1976;
\markcite{her86}Herbig \& Terndrup 1986). These values of column
density are comparable and are believed to be on the order of $10^{22}$
cm$^{-2}$ for the predominantly molecular clouds in the Galaxy
(\markcite{sol87}Solomon \etal\ 1987; \markcite{elm89}Elmegreen 1989;
\markcite{mcl96}McLaughlin \& Pudritz 1996). In starbursts, the gas
density may vary depending on how long ago the burst of star formation
began. Initially, the collapse of the molecular clouds could give a
higher than average gas density. The density should decrease as the gas
is used up, new hot stars heat and expand the gas, and stellar winds
blow gas out of the galaxy. This heating and expansion of the gas in a
molecular cloud will decrease the density while leaving the column
density unchanged (\markcite{mcl96}McLaughlin \& Pudritz 1996). But,
increasing the temperature excites more molecules to higher rotational
levels, which decreases the opacity in the lower rotational levels. We
should then expect to see optically thick gas in gravitationally
sub-critical molecular clouds unless the clouds are heated sufficiently
by external sources such as the active star forming regions we see in
NGC 6822.

 This heating effect may be amplified in the low metallicity
environment of NGC 6822. Lower metallicity suggests a lower dust
content, which would reduce the amount of shielding by dust from UV
radiation from the star forming regions. With this reduced shielding,
the UV radiation is able to heat the molecular gas more efficiently
than is possible in higher metallicity (higher dust content)
environments, which could contribute to the reduced optical depth seen
in this galaxy. We might also expect optically thin gas in regions
where there is simply less molecular gas. There is evidence for a
deficit of molecular gas in the weak CO line strengths of NGC 6822. We
thus attribute the low optical depth of the molecular gas in NGC 6822
to the low molecular gas content in this metal-poor environment. Our
data are insufficient to sort out the extent to which each factor
(namely low gas content and low metallicity) contribute to the low
optical depth.

Optically thick gas (assuming LTE) can only produce line ratios $\leq$
1 whereas optically thin gas can produce \twcott/\jto\ line ratios of
$\gtrsim$ 2.5 for temperatures greater than 100 K
(\markcite{wal91}Walker 1991). It is possible that only a fraction of
the gas in MC 2 is optically thin. For example, the cloud
could be a mixture of 10\% optically thin gas and 90\% optically thick
gas. Assuming the optically thin CO emission is not blocked by the
optically thick emission, we would then expect our line ratios to be on
the order of 1$\times$0.9 + 2.5$\times$0.1 = 1.15 for a 100 K cloud. A
rigorous calculation of the effect of mixtures of optically thin and
thick gas is complicated and beyond the scope of this paper. However,
this simple calculation suggests that MC 2 contains at least
$\sim$25\% optically thin gas. For simplicity, we will assume that the
observed high \twcott/\jto\ line ratio indicates that the entire cloud
is optically thin.

With only one line ratio for MC 2 (\twcott/\jto), we can only
place crude limits on the temperature and density. However, we do not
need to make any assumptions about the abundance ratio, since our data do
not include any \thco\ observations. Figure \ref{N6822lvg} shows a
possible solution for a temperature of 100 K. This temperature seems
reasonable, considering MC 2 is located within a massive star
forming region. If we assume a temperature any lower than 50 K, we would
require densities greater than 10$^6$ cm$^{-3}$ to achieve the observed
line ratios. While this is not impossible, it seems more likely that this
line ratio is a result of higher temperatures. For a temperature of 100
K, we can put a lower limit of 10$^{4.2}$ cm$^{-3}$ on density and an
upper limit of $10^{17}$ cm$^{-2}$ (km s$^{-1})^{-1}$ on column density
per unit velocity. The full-width half maximum velocity of the emission
lines in NGC 6822 is 5.4 $\pm$ 0.9 km s$^{-1}$ which implies a true upper
limit on column density of $5 \times 10^{17}$ cm$^{-2}$. 

\placefigure{N6822lvg}

If we increase the temperature above 100 K, we see that lower H$_2$
densities are allowed by the models. The observed line ratios can be
produced with H$_2$ densities as low as 100 cm$^{-3}$ with temperatures
of 200 K, but the CO column densities increase to $> 5 \times 10^{18}$
cm$^{-2}$. Assuming a number density ratio of {\it n}(CO)/{\it
n}(H$_2$) of 10$^{-4}$ (\eg\ \markcite{gen92}Genzel 1992) we obtain an
H$_2$ column density of about $5 \times 10^{22}$ cm$^{-2}$. This column
density is uncomfortably close to the critical column density for
molecular cloud collapse of $\sim 10^{22}$ cm$^{-2}$. As it seems
unlikely that we are fortunate enough to observe a cloud that is on the
brink of collapse, we will use the above value of the CO column density
of $5 \times 10^{17}$ cm$^{-2}$ in further calculations.

\subsection{Molecular Masses \label{mass}}

The intensity of the CO emission can be related to the molecular mass
of a cloud using the equation \begin{equation}
M_{\rm mol} = 1.61 \times 10^4~\left({\alpha\over{\alpha_{\rm Gal}}}
\right)\left ({115~{\rm GHz}\over{\nu}}\right )^2 ~ d^2_{\rm Mpc} ~
{S_{\rm CO} \over{R}} ~ \rm M_{\odot} \label{eqM} \end{equation}
(\markcite{wil90}Wilson \& Scoville 1990; \markcite{wil95}Wilson 1995)
where $S_{\rm CO}$ is the \twcoto\ flux in Jy km s$^{-1}$, $R$ is the
\twcoto/\joz\ line ratio, $\nu$ is the frequency of the emission (230
GHz for the \jto\ transition), $d_{\rm Mpc}$ is the distance to the
cloud in Mpc, $\alpha$ is the CO-to-H$_2$ conversion factor for the
galaxy, and $\alpha_{\rm Gal}$ is the Galactic value ($3\pm 1 \times
10^{20}$ cm$^{-2}$(K km s$^{-1}$)$^{-1}$, \markcite{str88}Strong
\etal\ 1988; \markcite{sco87}Scoville \& Sanders 1987). The IC 10
\twcoto/\joz\ line ratio was calculated by combining
\markcite{bec90}Becker's (1990) IRAM \twcooz\ (21$''$ beam) with our
JCMT \twcoto\ data (22$''$ beam)(see Table \ref{ratios}). For NGC 6822,
a \twcoto/\joz\ ratio of 0.7 is assumed. The CO-to-H$_2$ conversion
factor ($\alpha$) is a globally averaged property of the galaxy and
hence there are uncertainties involved in its use in one specific
region of the galaxy. Recent studies have suggested that the global
CO-to-H$_2$ conversion factor in IC 10 may be much higher than in our
own Galaxy ($\alpha/\alpha_{\rm Gal}$ = 40 -- 100, \markcite{isr97}Israel 1997; \markcite{mad97}Madden
\etal\ 1997). However, the conversion factor has been directly measured
on smaller scales for the regions we are studying by Wilson (1995), and
so we adopt her values of $\alpha/\alpha_{\rm Gal} = 2.7 \pm 0.5$ for
IC 10 and $2.2 \pm 0.8$ for NGC 6822 and note that this gives the lower
limit on the mass of the molecular gas. We use 34.6 Jy K$^{-1}$
($\eta_{\rm ap}$ = 0.45) and 29.4 Jy K$^{-1}$ ($\eta_{\rm ap}$= 0.53)
to convert the JCMT data from Kelvins ($T_{\rm A}^\star$) to Janskys in IC
10 and NGC 6822 respectively (\markcite{kra86}Kraus 1986). We assume a
coupling efficiency ($\eta_{\rm c}$) of 0.7 to correct our observed
fluxes to true fluxes. The virial masses from \markcite{bec90}Becker
(1990) were corrected to the new IC 10 distance of 0.82 Mpc. The
CO-to-H$_2$ conversion factor is only accurate to within $\sim$ 30\%
while our fluxes are typically accurate to about 10\%. We therefore
adopt a total uncertainty of 40\%. The results are shown in Table
\ref{masses}.

\placetable{masses}

It is interesting that the masses calculated here for the IC 10 regions
are larger than the virial masses ($M_{\rm vir}$) found by
\markcite{bec90}Becker (1990). Our beam size of 22$''$ covers a
diameter of 88 pc at the distance of IC 10. Although this beam is
smaller than the dimensions of the clouds in IC 10 as measured by
Becker using a comparable ($21''$) beam size (\eg\ IC 10:MC 1/2
measured 107 pc $\times$ 98 pc), the IRAM dimensions are not much
larger than the beam and were not deconvolved from the beam. Thus our
single position measurements probably detect most of the emission in
each region. Given the large uncertainties, the molecular mass of all
the clouds in IC 10 lie within 2$\sigma$ of the virial masses
calculated by \markcite{bec90}Becker (1990). If the clouds were not
gravitationally bound, the virial masses would be larger than the
molecular masses. We thus conclude that these quite large structures
are gravitationally bound.

The mass of IC 10:MC 1/2 ($\sim 5 \times 10^6$ M$_{\odot}$) indicates
that it may be either a large giant molecular cloud (GMC, $M$ = $10^5$
-- $10^6$ M$_{\odot}$, \markcite{san85}Sanders; Scoville, \& Solomon
1985) or a small giant molecular association (GMA, $M$ = $10^7$ --
$10^8$ M$_{\odot}$ \markcite{ran90}Rand \& Kulkarni 1990). For star
formation to occur, we would naturally expect that the molecular clouds
in which the stars form must be gravitationally bound. All of the GMCs
observed in IC 10 appear to be gravitationally bound
(\markcite{wil91}Wilson \& Reid 1991; \markcite{wil95}Wilson 1995). The
larger region IC 10:MC 1/2 also appears to be gravitationally bound.
Small GMAs seen in M33 all appear unbound by factors of about five {\it
except} in giant \ion{H}{2} regions (\markcite{wil89}Wilson \& Scoville
1989; \markcite{wil92}1992). In M51, the star formation efficiency is
enhanced in the spiral arms where GMAs are seen (\markcite{ran90}Rand
\& Kulkarni 1990). Individual GMCs inside the potential wells of the
bound GMAs may be colliding at a much higher rate than isolated GMCs,
which could increase the star formation rate and the star formation
efficiency. Thus the presence of large bound molecular structures in IC
10 may provide a possible explanation for the high star formation rate
in IC 10.
 
For NGC 6822:MC 2, the molecular mass is a factor of 3.4 larger than
the virial mass calculated by \markcite{wil94}Wilson (1994) using the
Owens Valley Millimeter-Wave Interferometer. This large mass
measurement may be caused by the larger 22$''$ beam overlapping
portions of the nearby cloud (NGC 6822:MC 1), which is approximately
17$''$ away. Also, the virial masses for NGC 6822 are taken from
interferometer data, which are insensitive to diffuse structures that
can be detected by single dishes such as the JCMT. Finally, in
determining the molecular mass for NGC 6822:MC 2, we assumed a
\twcoto/\joz\ value of 0.7. Given the rather high \twcott/\jto\ ratio,
we might expect this cloud to have an unusually high
\twcoto/\joz\ ratio as well. Increasing the \twcoto/\joz\ line ratio to
a value of $\sim$1.0 would give molecular and virial mass estimates
that agree within the uncertainties.

The CO-to-H$_2$ conversion factor relative to the Galactic value is
determined by assuming the clouds are virialized and setting the
molecular mass $M_{\rm mol}$ from equation (\ref{eqM}) equal to the
virial mass $M_{\rm vir}$. This method gives the relation $M_{\rm
vir}/M_{\rm mol} = \alpha/\alpha_{\rm Gal}$ (\eg\ \markcite{wil95}Wilson
1995). Equation (\ref{eqM}) assumes that the gas is optically thick and
emission comes from the surface. However, MC 2 in NGC 6822 appears to be
optically thin. For an optically thin cloud, the emission from the entire
volume of the cloud can escape and we can get a much higher flux. This
effect would result in an overestimate of the molecular mass found
using equation (\ref{eqM}), which in turn would result in an
underestimate of $\alpha/\alpha_{\rm Gal}$. Even if only part of the
cloud were optically thin, we would still be overestimating molecular
masses. Because the clouds in NGC 6822 were not fully resolved, the
derived virial masses were only upper limits and hence the value of
$\alpha/\alpha_{\rm Gal}$ derived by \markcite{wil94}Wilson (1994) is
also an upper limit. Since NGC 6822:MC 2 is optically thin, we now see
that $M_{\rm mol}$ was also overestimated, which would work to balance
the previous uncertainty and make the value of $\alpha/\alpha_{\rm
Gal}$ found by \markcite{wil94}Wilson (1994) more reliable. However,
since we do not know the extent of each mass overestimate, we cannot
make precise corrections to the previously published values of $\alpha$.

Given that the \twco\ emission from the cloud NGC 6822:MC 2 is
optically thin, we have a rare opportunity to calculate the mass of the
cloud directly. Using the upper limit on column density determined from
the LVG analysis, we can estimate the mass by adopting a number density
ratio of {\it n}(CO)/{\it n}(H$_2$) = 10$^{-4}$ (\markcite{gen92}Genzel
1992) and multiplying by the area of the cloud $\pi ab$, where $a$ and
$b$ are the major and minor axes of the elliptical cloud. It is
probably most appropriate to use the dimensions of the cloud obtained
in the interferometry maps ($36 \times <13$ pc, \markcite{wil94}Wilson
(1994)) instead of the JCMT beam size. The LVG model indicates that the
CO column density can be at most $5.3 \times 10^{17}$ cm$^{-2}$ (at
temperatures of 100 K) to produce the observed line ratio, which leads
to a total mass of H$_2$ of $< 1.2 \times 10^5$ M$_{\odot}$. This
mass is in good agreement with the molecular mass derived using
equation (\ref{eqM}), and agrees with the virial mass to within a
factor of two. As discussed in \S\ref{mass} we expect the mass derived
from single dish flux to be overestimated. Since the virial mass is
comparable to the mass derived from the column density, it is likely
that the molecular cloud MC 2 in NGC 6822 is gravitationally bound.

\section{Implications for the CO-to-H$_2$ Conversion Factor \label{alpha}}

It has been proposed that the CO-to-H$_2$ conversion factor ($\alpha$) is
not only dependent on density and brightness temperature ($\propto
\sqrt{\rho}/T_{\rm b}$) but also on the metallicity (\markcite{dic86}Dickman,
Snell, \& Schloerb 1986; \markcite{mal88}Maloney \& Black 1988;
\markcite{sak96}Sakamoto 1996). It is possible that the lower metallicity
in irregular galaxies may change the typical temperatures and densities
of the molecular clouds they contain. Although this effect may confuse
any observed relation between $\alpha$ and metallicity, in essence it
will simply introduce inter-relationships between $\rho$, $T_{\rm b}$ and
metallicity. Overall, the expression for $\alpha$ will still be a function
of metallicity.

The line ratios for the clouds in IC 10 agree with those found in M33 by
\markcite{wil97}Wilson \etal\ (1997) within the measurement uncertainties
(see Table \ref{ratios}). If we assume that the line widths are the same
for both transitions, then the line ratios are essentially ratios of
radiation temperature $T_{\rm R}$ (\eg\ \markcite{wal91}Walker 1991). The
radiation temperature is related to the cloud excitation temperature and
density by
\begin{equation}
 T_{\rm R} = \left [J_{\nu}(T_{\rm ex}) - J_{\nu}(T_{\rm bg}) \right ](1 - e^{-\tau}) \label{eqT}
\end{equation}
where $J_{\nu}(T)$ is the Planck function, $T_{\rm bg}$ is the background
temperature, and $T_{\rm ex}$ is the excitation temperature of the cloud.
The optical depth of the cloud is given by
\begin{equation}
\tau = \int_0^D K \rho ~dr \label{eqt}
\end{equation}
where $D$ is the depth of the cloud, $\rho$ is the density profile of the
cloud and $K$ is the absorption coefficient. For the optically thin case
where $\tau \ll 1$, the \twcott/\jto\ line ratio becomes
\begin{equation}
{T_{\rm R}^{3-2} \over{T_{\rm R}^{2-1}} } = { f^{3-2}\left [J_{\nu}(T_{\rm
ex}^{3-2}) - J_{\nu}(T_{\rm bg}) \right ]\tau^{3-2} \over{f^{2-1}\left
[J_{\nu}(T_{\rm ex}^{2-1}) - J_{\nu}(T_{\rm bg}) \right ]\tau^{2-1} } }
\end{equation}
where $f$ is the beam filling factor of the emission and the superscript
3-2 indicates those values pertaining to the \jtt\ transition. For the
optically thick case, $\tau \gg 1$, the line ratio has the form
\begin{equation}
{T_{\rm R}^{3-2} \over{T_{\rm R}^{2-1}} } = { f^{3-2}\left [J_{\nu}(T_{\rm
ex}^{3-2}) - J_{\nu}(T_{\rm bg}) \right ] \over{f^{2-1}\left
[J_{\nu}(T_{\rm ex}^{2-1}) - J_{\nu}(T_{\rm bg}) \right ] } }
\end{equation}
Both these line ratios are a complicated function of the excitation
temperature. Thus to produce similar line ratios in clouds with different
physical conditions would require a pathological interdependence of
density and temperature to produce the same excitation
temperature for the low-$J$ CO transitions. We feel that the similarity
in the line ratios in the clouds in IC 10 and M33 indicates similar
physical conditions (namely similar $\rho$ and $T$), and thus the difference in
the measured value of $\alpha$ (\markcite{wil95}Wilson 1995) can be
attributed to the difference in metallicity of the spiral galaxy M33 and
the metal poor dwarf irregulars IC 10 and NGC 6822.

\section{Conclusions \label{concl}}

We have observed the \twcoto\ and \jtt\ lines at a few locations in the
dwarf irregular galaxies IC 10 and NGC 6822. In addition, we have
observed the \thcoto\ and \jtt\ lines for molecular clouds in IC 10.
The clouds are located in a variety of star formation environments. In
general, the clouds in IC 10 show line ratios similar to those found in
M33 by \markcite{wil97}Wilson \etal\ (1997). We find that the
\twcott/\jto\ ratio increases with increasing proximity to \ion{H}{2}
regions as previously found by \markcite{wil97}Wilson \etal\ (1997) in
M33. This result may indicate a general relation that holds true for
all galaxies (and galaxy types) and is likely the result of a
combination of extreme temperature and density conditions near
star-forming regions. In this low metallicity (and in NGC 6822, low CO column
density) environment, this effect is more conspicuous than it is in
the higher metallicity spiral galaxy M33.

Combining the JCMT line ratios in IC 10:MC 1/2 with large velocity
gradient models indicates that this region has a kinetic temperature of
100 K, densities of $10^4 - 10^5$ cm$^{-3}$ and a column density of $6
\times 10^{18}$ cm$^{-2}$. The [\twco]/[\thco] abundance is between 50
and 70. The inclusion of the IRAM data from \markcite{bec90}Becker (1990)
suggests that there are two distinct emitting regions, a high density
region where the upper $J$ transitions dominate, and a lower density
region where the lower $J$ transitions dominate. For IC 10:MC 6/7/8, if we
assume a temperature of 20 K and [\twco]/[\thco] = 50, we obtain
densities of $10^{1.6} - 10^4$ cm$^{-3}$ and column densities of $2
\times 10^{17} - 2 \times 10^{19}$ cm$^{-2}$.
 
We find an unusually high \twcott/\jto\ line ratio in NGC 6822 which we
attribute to optically thin CO emission resulting from the combination
of a low molecular gas content in a low metallicity environment. An LVG
analysis of the \twcott/\jto\ line ratio in NGC 6822:MC 2 indicates the
presence of hot (T $\geq$ 100 K), optically thin gas. The density of
this cloud is likely greater than $10^{4.2}$ cm$^{-3}$ and its column
density is less than $5 \times 10^{17}$ cm$^{-2}$.

The optically thin CO emission in NGC 6822 allows a rare opportunity to
calculate the mass of molecular gas directly. Mass calculations from
two separate techniques determine the molecular mass to be $1.2 -
2.1\times 10^5$ M$_\odot$. This agrees with the virial mass to within a
factor of $\sim$3. Flux measurements yield masses for the cloud
complexes in IC 10 which indicate that these regions are
gravitationally bound. The masses of the clouds range from $0.83 - 5.2
\times 10^6$ M$_{\odot}$. These regions may represent either large
giant molecular clouds or small giant molecular associations. The
presence of these large gravitationally bound structures may help
explain the high star formation rate in IC 10.

The similarity in the line ratios between IC 10 and M33 is most easily
understood if the physical conditions in these clouds are also similar,
in which case any variation in the CO-to-H$_2$ conversion factor between
these two galaxies can be attributed to differences in their metallicity.
The increase in the \twcott/\jto\ ratio near star forming regions may be a
result of either higher temperatures, higher densities or some
combination of the two. Upcoming SCUBA observations will provide
temperature measurements which will enable us to better determine the
physical processes at work.

\acknowledgments The JCMT is operated by the Royal Observatories on
behalf of the Particle Physics and Astronomy Research Council of the
United Kingdom, the Netherlands Organization for Scientific Research, and
the National Research Council of Canada. This research has been supported
by a research grant to C. D. W. from NSERC (Canada).

\clearpage

\figcaption[f1.ps]{The \twcoto\ (solid line), \thcoto\ (dotted
line), and \twcott\ (dashed line) spectra observed in 3 regions in IC 10
and NGC 6822. The \jtt\ data have been convolved to match the $22''$ beam
of the \jto\ data. The \thcoto\ line has been scaled up by a factor of
three.
\label{spectra}}

\figcaption[f2.ps]{The \twcott\ (dashed line) and \thcott\ (solid
line) spectra observed towards the region MC 1/2 in IC 10. The spectra were
obtained with a $15''$ beam. The \thcott\ spectrum is scaled up by a
factor of three. 
\label{MC12spectra}}

\figcaption[f3.ps]{Channel by channel line ratios (in $T_{\rm MB}$)
for the 3 regions observed. The solid line represents the \twcott/\jto\
line ratio (scaled up by a factor of 10), the dashed line represents the
\twco/\thcott\ line ratio, and the dotted line represents the
\twco/\thcoto\ line ratio. Only those channels with a signal to noise
ratio greater than two are shown. 
\label{3ratios}}

\figcaption[f4.ps]{Solutions for the density and column density for the
region IC 10:MC 1/2 using a large velocity gradient code. This figure
shows the solution for the three JCMT line ratios at a kinetic
temperature of 100 K and a ${\rm [^{12}CO]/[^{13}CO]}$ abundance of
50. The solid lines indicate the 1$\sigma$ limits of the
\twco/\thcott\ line ratio of $5.3\pm0.4$. The dotted lines represent
the \twcott/\jto\ line ratio of $0.9\pm0.11$ and the dashed lines
represent the \twco/\thcoto\ line ratio of $8.2\pm0.6$. Also included
is the \twcoto/\joz\ line ratio of $0.64\pm0.05$ (\joz\ data from IRAM,
Becker 1990), shown as the dot-dashed lines. Solutions are indicated by
the shaded regions where the line ratios overlap. In this case there
are three separate regions where three of the four line ratios overlap
which may indicate two or more distinct emission regions within the
clouds.
\label{MC12lvg}}

\figcaption[f5.ps]{Typical LVG solution for the region MC 6/7/8 in IC
10 for a kinetic temperature of 20 K and a ${\rm [^{12}CO]/[^{13}CO]}$
abundance of 50. The dotted lines represent the 1$\sigma$ limits of the
\twcott/\jto\ line ratio of $0.66\pm0.11$ while the dashed lines indicate
the \twco/\thcoto\ line ratio of $8.3\pm1.4$. The shaded region indicates
the solutions for the JCMT data. Also included is the \twcoto/\joz\ line
ratio of $0.50\pm0.06$ (\twcooz\ data from Becker 1990). The \twcoto/\joz\
line ratio does not overlap with either of the JCMT line ratios. 
\label{MC68lvg}}

\figcaption[f6.ps]{This figure shows one possible temperature
(100 K) solution for the region MC 2 in NGC 6822. The dotted lines show the
1$\sigma$ limits of the \twcott/\jto\ line ratio of $1.4\pm0.07$. With
only a single line ratio, it is only possible to place lower limits on the
temperature ($> 50$ K) and density ($> 10^4$ cm$^{-3}$). 
\label{N6822lvg}}


\begin{deluxetable}{lcccccc}
\tablecaption{CO Line Ratios of Molecular Clouds in IC 10 and NGC 6822  \label{ratios}}
\tablewidth{0pt}
\tablehead{  \colhead{Cloud\tablenotemark{a}}
            &\colhead{$\alpha(1950)$}
            &\colhead{$\delta(1950)$} 
            &\colhead{$\frac{{\rm ^{12}CO~J=2-1}}{{\rm ^{13}CO~J=2-1}} $} 
            &\colhead{$\frac{{\rm ^{12}CO~J=3-2}}{{\rm ^{12}CO~J=2-1}} $} 
            &\colhead{$\frac{{\rm ^{12}CO~J=3-2}}{{\rm ^{13}CO~J=3-2}} $}
            &\colhead{$\frac{{\rm ^{12}CO~J=2-1}}{{\rm ^{12}CO~J=1-0}} $}
          }
\startdata
IC 10:MC 1/2 & 00:17:46.2 &  +59:00:28 & 8.2 $\pm$  0.6  & 0.90 $\pm$ 0.11 & 5.3 $\pm$ 0.4 & 0.64 $\pm$ 0.05 \nl
IC 10:R2 & 00:17:34.7 & +59:02:05 & ... & ... &  ... & 1.18 $\pm$ 0.1 \nl
IC 10:MC 4 &00:17:37.5 & +59:04:08 & ... & ... & ... & 0.65 $\pm$ 0.06\nl
IC 10:MC 6/7/8 & 00:17:40.1& +59:04:38 & 8.3 $\pm$  1.4  & 0.66 $\pm$ 0.11 & ... & 0.50 $\pm$ 0.06 \nl
NGC 6822:MC 2 & 19:42:03.3 & --14:50:27 &  ...  & 1.54 $\pm$ 0.08 & ...& ... \nl 
\noalign{\vskip 1.0ex}
\tableline
\noalign{\vskip 1.0ex}
M33 average\tablenotemark{b} & ... & ... &7.2 $\pm$ 1.1  & 0.69 $\pm$ 0.15 & ... & 0.67 $\pm$ 0.19 \nl
\enddata
\tablenotetext{~}{All data obtained at JCMT 15 m, except for the \twcooz\
data which are taken from \markcite{bec90}Becker (1990). All line ratios
refer to a $22''$ beam, except for the \twco/\thcott\ ratio which is for
a $15''$ beam. The line ratios are the average of all
channels with a signal to noise greater than two (see Figure \ref{3ratios})
and the uncertainties are the standard deviation of the mean. }
\tablenotetext{a}{Cloud designations from \markcite{wil91}Wilson \& Reid 1991,
\markcite{wil94}Wilson 1994, \markcite{wil95}Wilson 1995 except R2
(\markcite{bec90}Becker 1990).}
\tablenotetext{b}{From \markcite{wil97}Wilson \etal\ 1997, \markcite{tho94}Thornley \& Wilson 1994. }

\end{deluxetable}


\begin{deluxetable}{lccc}
\tablecaption{Masses of Molecular Clouds in IC 10 and NGC 6822  \label{masses}}
\tablewidth{0pt}
\tablehead{\colhead{Cloud} & \colhead{$S_{\rm CO (J=2-1)}$ } & \colhead{Molecular Mass\tablenotemark{a} } & \colhead{Virial Mass\tablenotemark{b}} \nl
\colhead{~} & \colhead{(K km s$^{-1}$)($T_{\rm A}^{\star}$)} & \colhead{(M$_{\odot}$)} & \colhead{(M$_{\odot}$)} }
\startdata
IC 10:MC 1/2 & 9.16 & $5.2 \times 10^6$  & $2.5 \times 10^6$  \nl
IC 10:R2 & 2.67 & $0.83 \times 10^6$  & $0.35 \times 10^6$   \nl
IC 10:MC 4 & 2.37 & $1.3 \times 10^6$  & $0.54 \times 10^6$  \nl
IC 10:MC 6/7/8 & 2.35 & $1.7 \times 10^6$  & $0.66 \times 10^6$  \nl
NGC 6822:MC 2\tablenotemark{c} & 1.66 & $0.21 \times 10^6$ & $< 6.3 \times 10^4$ \nl
\enddata
\tablenotetext{a}{A CO-to-H$_2$ conversion factor ($\alpha/\alpha_{\rm Gal}$) of
2.7 and 2.2 has been assumed for IC 10 and NGC 6822 respectively.}
\tablenotetext{b}{Virial masses taken from \markcite{bec90}Becker (1990) except NGC 6822:MC 2 (\markcite{wil94}Wilson 1994).}
\tablenotetext{c}{A \twcoto/\joz\ ratio of 0.7 has been assumed.}
\end{deluxetable}


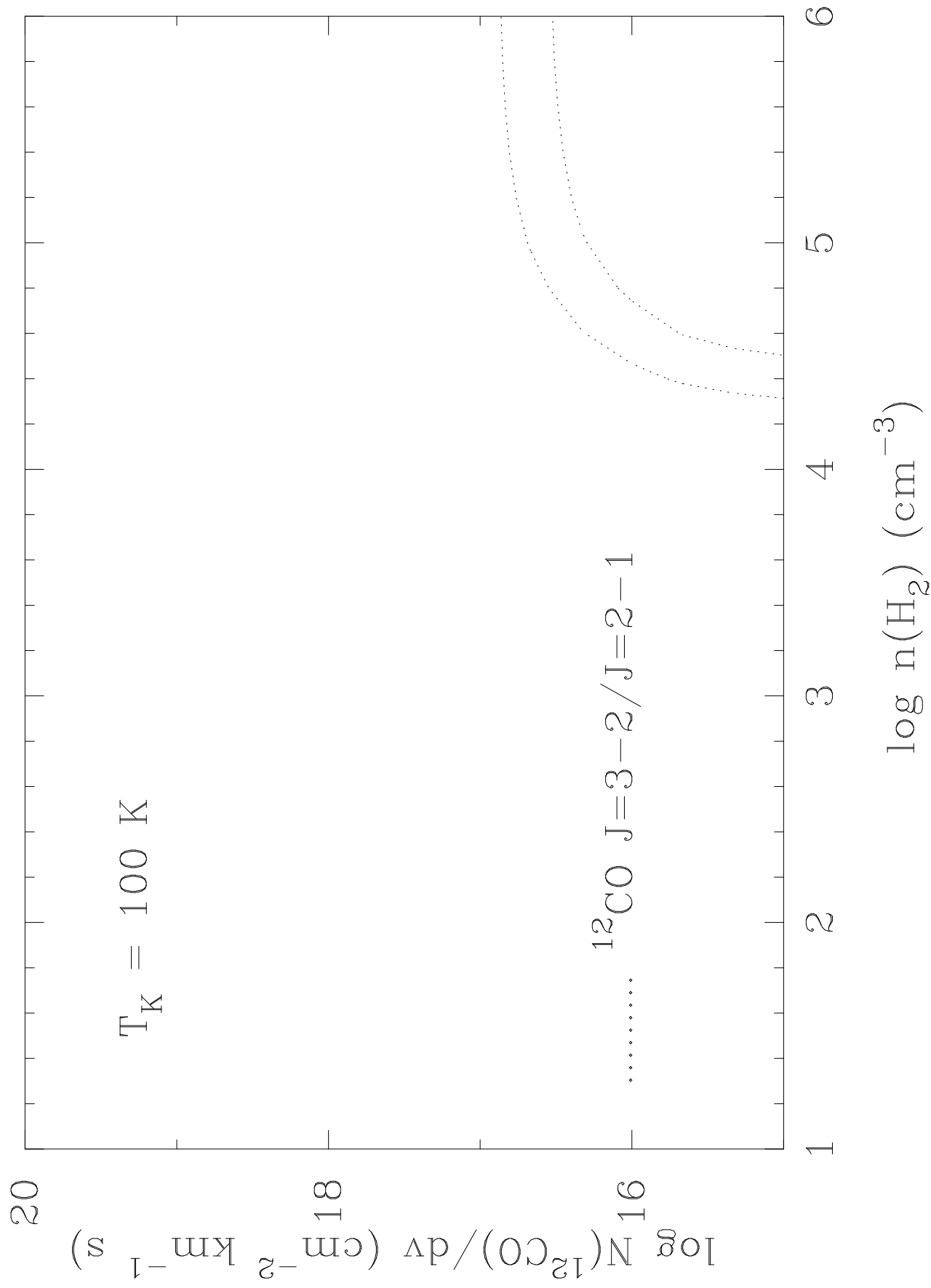
\end{document}